\documentclass[manuscript=article, journal=jpcafh]{achemso}
\setkeys{acs}{maxauthors=99}
\setkeys{acs}{etalmode=truncate}
\usepackage{graphicx}
\usepackage[font={small}]{caption} 
\usepackage{amsmath, amssymb}
\usepackage{float}
\usepackage{xr}
\usepackage{placeins}
\usepackage{adjustbox}

\usepackage{xcolor}
\usepackage{soul}
\usepackage[colorlinks=true,
            linkcolor=black,
            urlcolor=black,
            citecolor=black]{hyperref}

\newcommand*{\blauw}[1]{#1}

\SectionNumbersOn

\author{Soumajit Dutta}
\affiliation[University of Chicago]
{Pritzker School of Molecular Engineering, University of Chicago, Chicago, Illinois 60637, United States of America}
\author{Cunzhi Zhang}
\affiliation[University of Chicago]
{Pritzker School of Molecular Engineering, University of Chicago, Chicago, Illinois 60637, United States of America}
\author{Gustavo Perez Lemus}
\affiliation[University of Chicago]
{Pritzker School of Molecular Engineering, University of Chicago, Chicago, Illinois 60637, United States of America}
\author{Juan J. de Pablo}
\affiliation[NYU]
{Tandon School of Engineering, New York University, New York, New York 11201, United States of America}
\author{Francois Gygi}
\affiliation[University of California Davis]
{Department of Computer Science, University of California Davis, California 95616, United States of America}
\author{Giulia Galli}
\affiliation[University of Chicago]
{Pritzker School of Molecular Engineering, University of Chicago, Chicago, Illinois 60637, United States of America}
\alsoaffiliation[University of Chicago]
{Department of Chemistry, University of Chicago, Chicago, Illinois 60637, United States of America}
\alsoaffiliation[Argonne National Laboratory]
{Materials Science Division and Center for Molecular Engineering, Argonne National Laboratory, Lemont, Illinois 60439, United States of America}
\author{Andrew L. Ferguson}
\affiliation[University of Chicago]
{Pritzker School of Molecular Engineering, University of Chicago, Chicago, Illinois 60637, United States of America}
\alsoaffiliation[University of Chicago]
{Department of Chemistry, University of Chicago, Chicago, Illinois 60637, United States of America}
\email{andrewferguson@uchicago.edu}

\title{Characterizing Defect Dynamics in Silicon Carbide Using Symmetry-Adapted Collective Variables and Machine Learning Interatomic Potentials}

\begin{document}
\begin{abstract}
\noindent Silicon carbide (SiC) divacancies are attractive candidates for spin defect qubits possessing long coherence times and optical addressability. The high activation barriers associated with SiC defect formation and motion pose challenges for their study by first-principles molecular dynamics. In this work, we develop and deploy machine learning interatomic potentials (MLIPs) to accelerate defect dynamics simulations while retaining \textit{ab initio} accuracy. We employ an active learning strategy comprising symmetry-adapted collective variable discovery and enhanced sampling to compile configurationally diverse training data, calculation of energies and forces using density functional theory (DFT), and training of an E(3)-equivariant MLIP based on the Allegro model. The trained MLIP reproduces DFT-level accuracy in defect transition activation free energy barriers, enables the efficient and stable simulation of multi-defect 216-atom supercells, and permits an analysis of the temperature dependence of defect thermodynamic stability and formation/annihilation kinetics to propose an optimal annealing temperature to maximally stabilize VV divacancies.
\end{abstract}

\maketitle

\clearpage
\newpage


\section{Introduction}

Solid-state spin defects play a crucial role in advancing various quantum technologies, including quantum sensing, communication, and computing \cite{Weber2010, Awschalom2018}. Silicon carbide (SiC), a wide-bandgap semiconductor, has emerged as a promising host for such defects owing to its well-established growth, doping, and fabrication techniques \cite{Lohrmann2017}. Among these, the divacancy (VV) configuration, which forms by the interaction of a silicon vacancy (V$_{Si}$) and carbon vacancy (V$_{C}$) has garnered significant attention as a spin qubit due to its desirable characteristics, such as long coherence times, optical addressability, and high-fidelity readout \cite{Wolfowicz2021}.

\begin{figure}[!ht]
  \includegraphics[width=\textwidth]{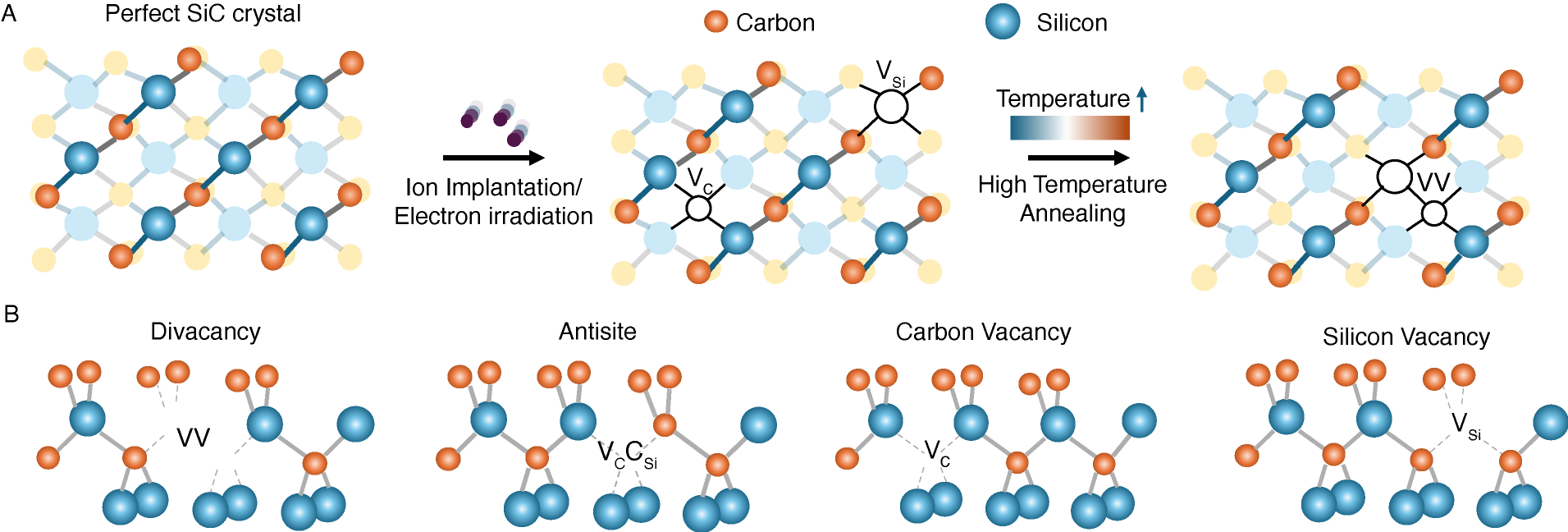}
  \caption{Vacancy defects in silicon carbide (SiC) crystals. (A) Schematic illustration of the divacancy formation process in SiC. Vacancies are typically created by ion implantation or electron irradiation followed by temperature annealing to promote high defect mobility and the formation of stable divacancies. (B) A number of vacancy defects commonly appear and interconvert during the annealing process including divacancies (VV), antisite vacancies (V$_{C}$V$_{Si}$), carbon vacancies (V$_{C}$), and silicon vacancies (V$_{Si}$).}
  \label{fgr:1}
\end{figure}

In SiC, VV defects are typically formed via ion implantation or electron irradiation followed by high-temperature annealing \cite{Wolfowicz2021} (Figure \ref{fgr:1}). Computational modeling offers a means to gain mechanistic understanding and insight into defect formation, transition, and migration, and can also predictively identify optimal thermal annealing protocols \cite{Carlsson2010, Son2006, Carlos2006, Son2007, Lee2021, Zhang2023, Kimura2025}. \textit{Ab initio} simulations have, however, reported activation barriers for typical defect transitions of several electronvolts \cite{Bockstedte2003, Rurali2003, Wang2013, Yan2020, Lee2021, Zhang2023, Zhang2024}. Yan \textit{et al.}\ performed \textit{ab initio} density functional theory (DFT) calculations using the PBE functional and reported migration barriers of 4.3 eV for V$_{Si}$ and 3.3 eV for V$_{C}$ in the 4H-SiC polytype \cite{Yan2020}. Zhang \textit{et al.}\ carried out calculations for the 3C-SiC polytype and obtained barriers of 4.05 eV for V$_{Si}$ and 3.87 eV for V$_{C}$ \cite{Zhang2024}. These high energy barriers make defect transitions a rare event and render unbiased \textit{ab initio} simulation of defect dynamics exceedingly computationally expensive. Simulation of defect dynamics using classical force fields is computationally efficient, but typically inaccurate due to the lack of explicit treatment of the important electronic effects \cite{Zhang2024}. 

Machine learning interatomic potentials (MLIPs) have created an opportunity to break this trade-off between efficiency and accuracy \cite{Unke2021}. MLIPs predict atomic forces and energies by learning many-body interactions typically using either kernel-based methods \cite{Bartk2010, Chmiela2018, Vandermause2020} or neural networks \cite{Smith2017}. The field has evolved rapidly since seminal work by Behler and Parrinello \cite{Behler2007} that introduced a symmetric, invariant representation of atomic local environments to predict system energies using neural networks. Since then, numerous more sophisticated machine learning models and atomic environment descriptors have been developed to capture atomic interactions with greater fidelity. For example, the kernel-based FLARE MLIP \cite{Vandermause2022}, built on the Atomic Cluster Expansion (ACE) descriptor \cite{Drautz2019}, was used to study pressure-induced phase transitions in SiC polytypes with accuracy comparable to \textit{ab initio} calculations and with high sampling efficiency \cite{Xie2023}. In recent years, equivariant neural network-based MLIPs have attracted growing interest due to their improved data efficiency and generalizability, enabled by the explicit encoding of physical symmetries in the hidden layers \cite{Batzner2022, schutt2021equivariantmessagepassingprediction, tholke2022torchmdnetequivarianttransformersneural, gasteiger2024gemnetuniversaldirectionalgraph, Wang2024}.

In this study, we employ the Allegro MLIP, which uses an E(3)-equivariant graph neural network architecture with atomic cluster expansion (ACE) descriptors \cite{Musaelian2023}. Allegro is selected among the variety of MLIP models for its attractive combination of high accuracy, data efficiency, and computational speed \cite{Nikidis2024, Leimeroth2025}. To ensure stability in Allegro-driven molecular dynamics simulations, we employ a prior potential to accelerate learning, improve extrapolation and constrain the MLIP from sampling unrealistic conformations \cite{Majewski2023}, and adopt an active learning approach to expand the training data to cover diverse thermally relevant system configurations. Using this protocol, we train an Allegro-based MLIP possessing \textit{ab initio} accuracy in just six rounds of active learning requiring DFT force and energy calculations for 1800 training configurations. The resulting MLIP reproduces five key defect transition activation free energies within the thermal fluctuation error (1 $k_bT$ $\approx$ 0.13 eV at 1500 K) of the DFT calculations and enables stable long-term simulations of a 216-atom supercell of SiC at an execution speed of approximately 6 ns/day on an NVIDIA L40S GPU. 
We deploy the trained MLIP to study defect transitions in 3C-SiC crystal polytype and study the temperature variation effect on divacancy formation from the dynamic interactions of two monovacancies. Our analyses of these long unbiased simulation trajectories reveals a subtle interplay of the relative stability and kinetic interconversion rates between competing defect structures and can inform materials processing strategies to maximally stabilize VV spin defects in SiC.


\section{Methods}

\subsection{SiC defect crystal}
All calculations in this work were conducted on a 3C-SiC polytype with the same 4.416~\AA\ at 1500 K lattice constant used in previous work \cite{Zhang2024}. Unless otherwise specified, a simulation supercell containing 216 atoms was used. For the multi-defect study, a larger 512-atom supercell was employed. Defects were introduced by selectively removing carbon and silicon atoms from the lattice. The Atomic Simulation Environment (ASE) Python package \cite{HjorthLarsen2017} was used for system construction and for converting between different file formats. 

\subsection{Active learning training of MLIP for SiC defect dynamics}
\begin{figure}[!ht]
  \includegraphics[width=\textwidth]{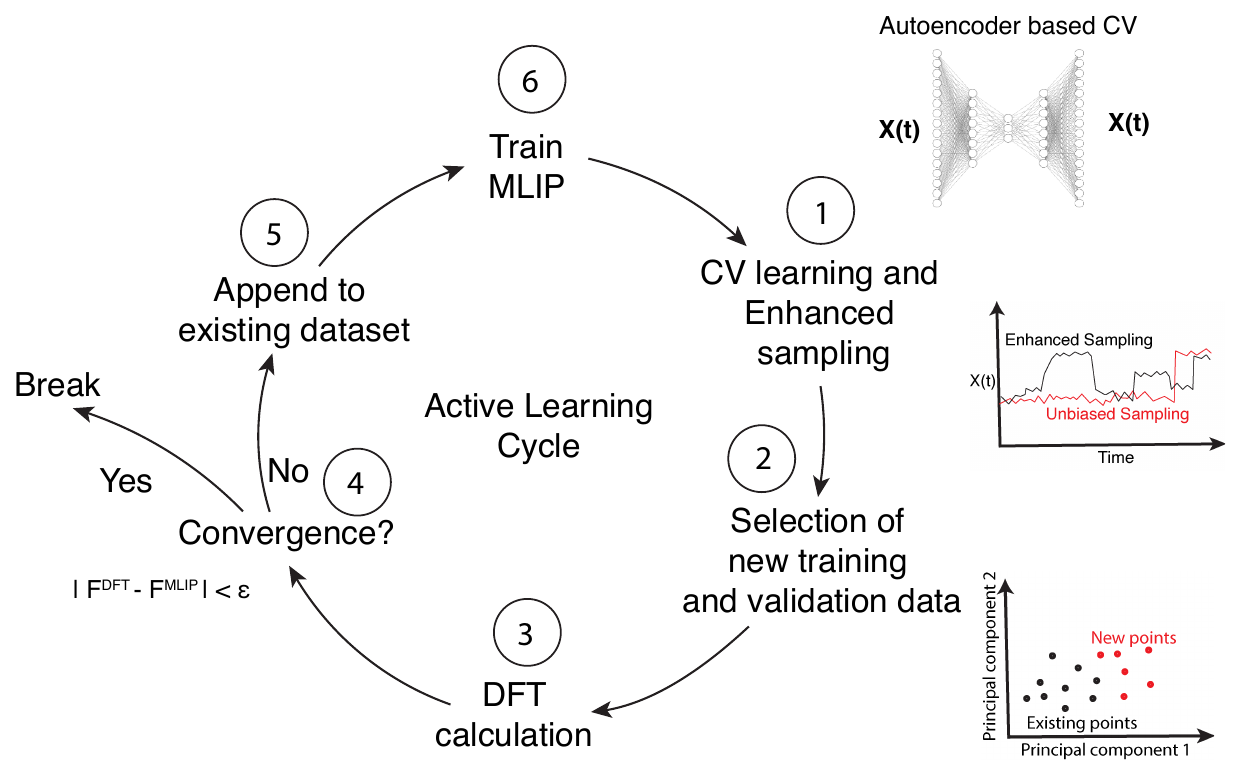}
  \caption{Active learning protocol for training an Allegro machine learning interatomic potentials (MLIP) for SiC. The cyclic process includes (1) enhanced sampling using an autoencoder-based collective variable (CV), (2) configuration selection, (3) density functional theory (DFT) evaluation of selected configurations, (4) convergence assessment, (5) training and validation data appending and (6) force field retraining with the updated dataset.}
  \label{fgr:2}
\end{figure}

Recent studies have highlighted the instability of long-time molecular dynamics simulations under MLIPs \cite{Wang2023, fu2023, Liu2023}. This drawback can ofter be attributed to the incorrect prediction of forces for conformations outside of training data due to limited extrapolation capabilities of the MLIP.  Therefore, the accuracy and stability of the MLIP relies on the ability to effectively capture a diverse range of local environments in the training data to enable accurate interpolation of the local environments encountered during simulations. To overcome the dual goals of generating diverse conformations and constructing a reliable force field using those conformations, we adopt an active learning approach that interleaves rounds of enhanced sampling of configurational space using the MLIP and retraining of the MLIP using the expanded set of training configurations (Figure \ref{fgr:2}). We describe here an overview of the six steps in the active learning workflow, and provide full technical details in the \blauw{Supplementary Methods} in the \blauw{Supporting Information}.

\textbf{Step 1: CV learning and enhanced sampling.} In the first step of the active learning cycle, we conducted enhanced sampling of configurational space using OPES-Explore, a variant of the On-the-fly Probability Enhanced Sampling (OPES) method, to efficiently explore diverse local defect configurations with the most recent Allegro MLIP trained in the prior round of active learning (\blauw{Figure S3}). Simulations were driven in three collective variables (CVs) learned over the enhanced sampling trajectories conducted in the previous active learning cycle using Permutationally Invariant Networks for Enhanced Sampling (PINES), designed to respect the underlying translational, rotational, and permutational symmetries of the crystal lattice \cite{Herringer2023}. PINES learns symmetry-adapted CVs by transforming a pairwise distance vector into a permutationally invariant vector (PIV) through block-wise sorting, then feeds this vector into a variational autoencoder (VAE). The PIV can be conceived as an atom-type specific pair correlation function that serves as a translationally, rotationally, and permutationally invariant featurization of an atomic configuration \cite{Pietrucci2018}, and the low-dimensional CVs extracted from the bottleneck layer of the VAE can be conceived as a low-dimensional distillation of the PIV \cite{Kingma_2019}. Importantly, the learned CVs preserve the same translational, rotational, and permutational symmetries of the PIV vector, which is desirable in improving the data efficiency of learning, and are also differentiable with respect to atomic coordinates, thus permitting their use within enhanced sampling calculations \cite{sidky2020machine, chen2018molecular, chen2018collective}.

Calculations were performed in a 216-atom supercell of the 3C-SiC polytype containing a V$_{C}$ and a V$_{Si}$ defect positioned within one coordination shell of each other. The atoms within the third coordination shell of the initial monovacancies were restrained to prevent the monovacancies from migrating too far apart and promote defect transitions and the exploration of diverse local defect environments. PIV vectors were constructed by considering all atoms residing within the two coordination shells around the defect, producing a length 435 vector. Atoms of the same element and coordination shell were treated as identical, yielding four permutationally indistinguishable atom-type groups that enabled the PINES CVs to distinguish defect reorientations while eliminating degeneracy from permutations of identical atoms. We employed a simple VAE architecture employing an encoder with hidden layers comprising 64, 32, and 16 neurons and a symmetric decoder. A bottleneck layer of 3 neurons was used to extract three learned CVs, which we found to be sufficient to resolve all of the important defect configurations and transitions. A custom PLUMED \cite{PLUMED2019} module implementing PINES is available in our GitHub repository at \url{https://github.com/Ferg-Lab/SiC_Allegro_MLIP.git}.

Molecular dynamics simulations were performed using the LAMMPS simulation package (v.~27 June 2024) \cite{Thompson2022}, employing GPU acceleration via the KOKKOS package \cite{johansson2025lammps}. All simulations were carried out in the canonical (NVT) ensemble using a Nos\'e-Hoover thermostat with a damping parameter of 0.1 ps to maintain the system temperature at 1500 K and employing a velocity Verlet integrator with a 1 fs time step \cite{allen2017computer}. Initial velocities were assigned from a Gaussian distribution with total momentum and angular momentum removed, and a 15 ps equilibration run conducted to relax the system. The neighbor list was updated every step with a bin size of 2.0 \AA, and dynamic neighbor checking was enabled. In each round of active learning, 10 ns of enhanced sampling were performed and trajectories written with a period of 0.1 ps. 

In the first round of active learning, we lack both an initial Allegro MLIP and learned PINES CVs. To ignite the active learning loop, we model the system using a pre-trained Fast Learning of Atomistic Rare Events (FLARE) MLIP developed for SiC crystals \cite{Xie2023} and drive enhanced sampling using PINES CVs learned from unbiased simulation trajectories performed using the classical Environment-Dependent Interatomic Potential (EDIP) \cite{Justo1998}. The EDIP potential is known not to provide a satisfactory description of defect dynamics \cite{Zhang2024}, but we use it to learn approximate initial CVs with which to initialize the first round of active learning (\blauw{Figure S1-2}). The pre-trained FLARE potential provided a reasonable starting model for exploring SiC crystal defect configurations (\blauw{Figure S4}), but due to the superior learning and data efficiency of equivariant graph neural network-based models \cite{Perego2024, Owen2024, xu2024molecular, perez2024importance, jin2024improving}, subsequent rounds of active learning are used to develop a SiC-specific Allegro-based MLIP. The FLARE and EDIP potentials were employed exclusively to initialize the first round of the active learning cycle and do not feature any further in the workflow.

\textbf{Step 2: Selection of training and validation data.} The enhanced sampling trajectories were analyzed to select new training and validation points for Allegro MLIP training and performance evaluation. Dissimilarity of configurations in the current enhanced sampling trajectories relative to the current set of training points were evaluated using a smooth overlap of atomic positions (SOAP) kernel \cite{Bartk2013}, which computes similarity by the dot product of average power spectrum vectors \cite{De2016} to present a translationally, rotationally and permutationally invariant representation of the local atomic environments obtained by expanding the local atomic density using orthogonal basis functions \cite{De2016}. Calculations were performed using the Python package DScribe \cite{Himanen2020}. The 300 most dissimilar configurations were added to the training set in each round of the active learning cycle. The validation set was defined by clustering all sampled configurations visited over the enhanced sampling trajectories collected to date into 100 cluster centers to provide a diverse sample of configurational space for evaluation of MLIP performance. The SOAP dissimilarity of points to the training data is positively correlated with the MLIP prediction error in the force relative to the DFT ground truth, providing support for a diversity maximizing strategy in curating the training data (\blauw{Figure S5}). 

\textbf{Step 3: DFT energy and force calculations.} Energy and force calculations for the selected training and validation conformations were performed using DFT with Qbox \cite{Gygi2008}. Calculations were conducted using the PBE functional \cite{Perdew1996} with optimized norm-conserving Vanderbilt pseudopotentials \cite{Schlipf2015} and a plane-wave kinetic energy cutoff of 55 Ry. The simulation supercell was assumed to be neutrally charged, and the calculations were conducted in a triplet spin state. Previous studies have shown that the activation temperature of defect formation depends on effective barriers that in turn depend on the charge state of the defect \cite{Zhang2024}. However the procedure adopted here to derive MLIPs requires the choice of a specific charge state. Our choice of a neutral state is an approximation which nevertheless allows us to understand trends on the stability of VVs as a function of temperature as well as kinetic effects, and to obtain valuable mechanistic insights. We demonstrate good agreement of our model predictions with experiment (\textit{vide infra}), but propose that future work may consider the development of charge- and spin-adaptable MLIPs in pursuit of superior physical realism. All DFT calculations were performed on single Intel Xeon E5-2680 CPUs. 

\textbf{Step 4: Convergence assessment.} The prediction accuracy of the Allegro MLIP relative to reference DFT forces was evaluated using a force-based convergence criterion. We seek to train a model with a force accuracy at the 1500 K simulation temperature of 1 $k_B T$/$\mathrm{\AA}$ $\approx$ 3 kcal/mol.$\mathrm{\AA}$ $\approx$ 0.13 eV/$\mathrm{\AA}$, which is in line with the typical performance expected from a MLIP for reliable molecular dynamics simulations and accurate property prediction \cite{Chmiela2018, Zuo2020}. Guided by this benchmark, we defined convergence of the active learning cycle as the point at which the Allegro MLIP achieves a mean absolute error (MAE) below 0.1 eV/$\mathrm{\AA}$ on both the training and validation datasets. When this threshold is satisfied, the active learning loop terminates; otherwise, additional rounds of sampling, DFT labeling, and model retraining are performed. 


\textbf{Step 5: Training and validation data update.} If the Allegro MLIP has not met the accuracy threshold on the validation data, we augment the existing training set with the 300 new training points and replace the existing validation set with the 100 new validation points spanning the configurational space. 

\textbf{Step 6: Allegro MLIP retraining.} The Allegro MLIP was retrained over the augmented training set. The Allegro model is implemented and trained using the NequIP v0.6.0 \cite{Batzner2022} and Allegro v0.2.0 \cite{Musaelian2023} Python packages and all calculations performed using a cutoff distance of 6~$\mathrm{\AA}$ for neighbor list construction conducted in double-precision floating point. The model returns pairwise energies, which are then appropriately summed to provide the per-atom and overall system energy. Atomic forces are computed by automatic differentiation of the predicted energies through the computational graph to guarantee a conservative force field. Training against the ground truth DFT energies and forces was conducted using a mean squared error (MSE) loss function with equal weighting of the energy and force terms, and the model weights and biases updated using the Adam optimizer \cite{kingma2017adammethodstochasticoptimization}. The model is exposed to all training points collected to date. The training set volume increases by 300 configurations in every pass through the active learning cycle, so later training rounds are exposed to comparatively more training data. To stabilize the force field in data-sparse regimes and prevent exploration of unphysical, high-energy configurations with atomic overlaps during enhanced sampling (\blauw{Figure S6}), we followed the approach adopted in several modern MLIP frameworks of employing a physically motivated prior potential \cite{Unke2021sp, Majewski2023, Jacobs2025}. Specifically, we employed a Lennard-Jones (LJ) prior and used a delta-learning strategy in which Allegro learns only the residual between the LJ prior and the reference DFT energies and forces. The parameters of the LJ prior were learned by linear least squares fitting against the ground truth DFT energies and an Allegro MLIP trained to fit the residuals between the LJ and DFT energies and forces (\blauw{Figure S7}). This approach furnished a simpler learning problem for the Allegro MLIP, stabilized the model against exploring thermally irrelevant high-energy regions of configurational phase space in the aggressive OPES-Explore enhanced sampling calculations (\blauw{Figure S8}) without any degradation in predictive accuracy (\blauw{Figure S9}). A custom implementation of the Allegro MLIP under the LJ prior potential and delta learning strategy is available in our GitHub repository at \url{https://github.com/Ferg-Lab/SiC_Allegro_MLIP.git}. All Allegro training was performed with GPU acceleration using NVIDIA’s CUDA framework on an NVIDIA L40S GPU.

\section{Results}

\subsection{Active learning training of Allegro MLIP for SiC defect dynamics} 

An active learning strategy was employed to train an Allegro MLIP for simulating defect dynamics in a 3C-SiC polytype with computational efficiencies approaching that of classical force fields ($\sim$6 ns/day on an NVIDIA L40S GPU) and \textit{ab initio} accuracy in force predictions ($<$0.1 eV/$\mathrm{\AA}$ mean average error). The E(3)-equivariant nature of the Allegro MLIP makes training inherently data efficient and by employing an active learning strategy combining data-driven learning of symmetry-adapted collective variables, enhanced sampling of the thermally-accessible phase space, a physically-motivated prior potential, and a diversity maximizing selection of training configurations, we converged the Allegro MLIP using just 1800 training configurations over the course of six active learning rounds. A total of $\sim$240 GPU-h were expended on conducting unbiased and enhanced sampling molecular simulations and $\sim$43 GPU-h on MLIP training using a NVIDIA L40S GPU and $\sim$72,000 CPU-h in computing ground truth energies and forces for the training and validation configurations on Intel Xeon E5-2680 CPUs.

\clearpage
\newpage
\begin{figure}[!ht]
  \includegraphics[width=1.0\textwidth]{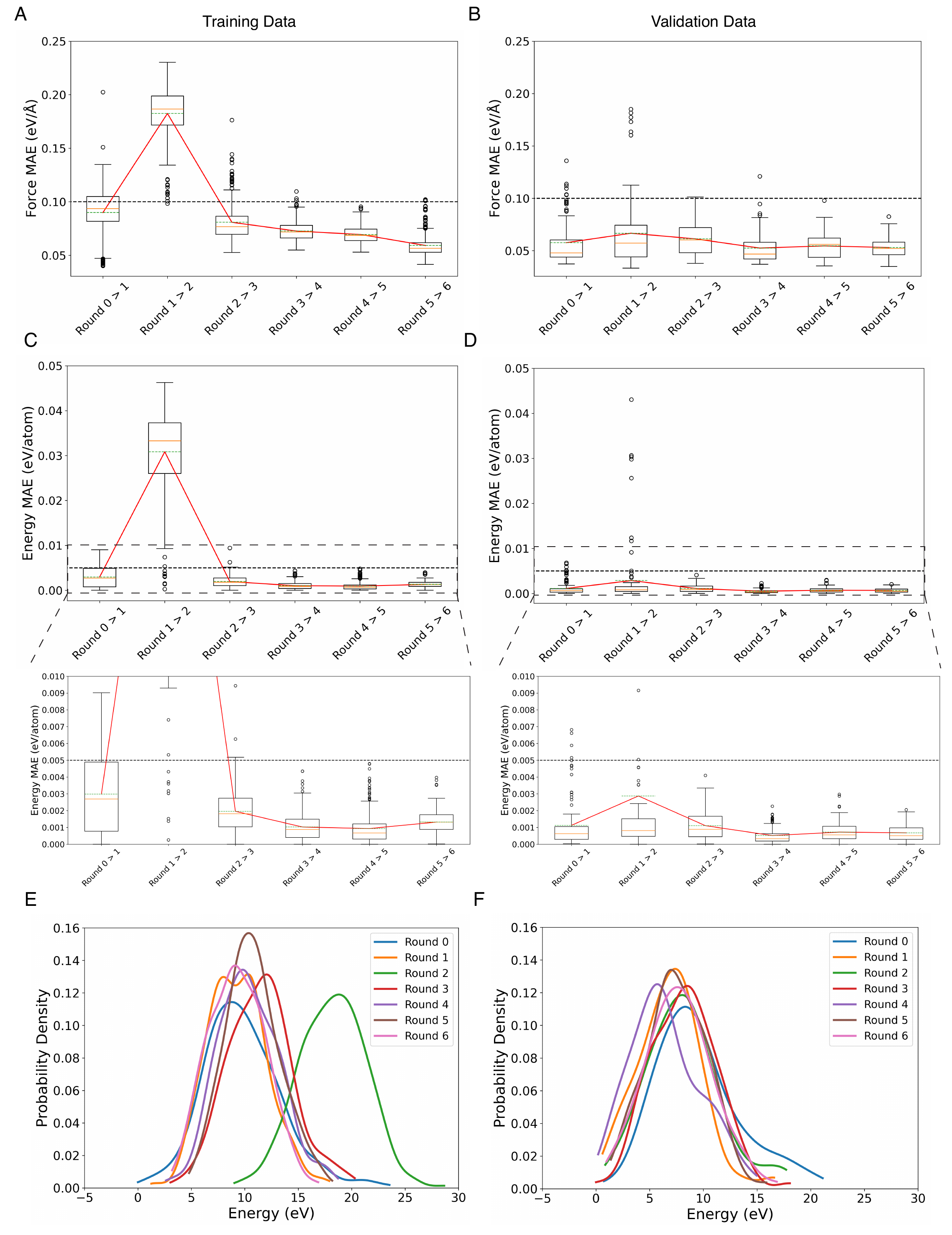}
\end{figure}
\clearpage
\newpage
\begin{figure}[!ht]
  \caption{Accuracy of the trained Allegro MLIP after each round of active learning. Box plots of the mean average error (MAE) in the force predictions relative to the DFT ground truth over (A) the 300 training configurations and (B) 100 validation configurations in each round of the active learning cycle. The green horizontal line within the box represents the mean of the distribution, and the orange horizontal line represents the median. The limits of the box correspond to the first (Q1) and third (Q3) quartiles, and the whiskers extend to the minimum and maximum values within 1.5 times the interquartile range (IQR) from the quartiles. Outliers in the distribution are explicitly shown. The red line connecting the means indicates the trend. The dotted horizontal line represents the desired force MAE threshold of 0.1 eV/\AA\ used to judge convergence of the model. (C,D) Analogous plots to panels (A,B), but illustrating the MAE in the per atom energy relative to the DFT ground truth. The dotted horizontal line represents an MAE in the energy of 0.005 eV/atom. Energy distribution of each round of (E) training and (F) validation configurations showing the diversity of configurations explored over each of the active learning rounds. The exploration of high-energy training configurations in Round 2 coincides with the spike in the MAE.} 
  \label{fgr:3}
\end{figure}

We present in Figure~\ref{fgr:3} the progression of the Allegro MLIP energy and force errors over the validation set at each round of the active learning cycle. The initial MLIP performs apparently quite well on the validation data collected in Round 1, with a mean average error (MAE) in the predicted forces relative to the DFT ground truth averaged over the training set of just 0.089 eV/\AA\ and over the validation set of just 0.057 eV/\AA, lying below the 0.1 eV/\AA\ threshold used to judge sufficiently accurate performance of the MLIP (Figure~\ref{fgr:3}A,B). The apparently good performance of the Round 1 model is, however, deceptive, since it is largely a result of quite low configurational diversity in the training and validation sets (Figure~\ref{fgr:3}E). The MLIP has not yet been challenged to train or predict on diverse defect configurations across the thermally accessible phase space. Furthermore, there is quite high variability in the force MAEs, with 35\% of the training configurations and 2\% of the validation configurations possessing force MAEs in excess of the 0.1 eV/\AA\ target threshold. In Round 2, the enhanced sampling calculations generate much higher configurational diversity in the training set (Figure~\ref{fgr:3}E), leading to an approximate doubling of the training set mean force MAE to 0.182 eV/\AA\ and an uptick in the validation force MAE to 0.067 eV/\AA. The diversity maximizing strategy to accumulating training configurations quickly pays dividends in subsequent active learning rounds, with the training mean force MAE plummeting to 0.081 eV/\AA\ and the validation force MAE to 0.061 eV/\AA\ in Round 3. The training force MAE monotonically decreases in all subsequent rounds while the validation force MAE plateaus in Round 4. We declare the MLIP converged in Round 6 where the training mean force MAEs meets the validation mean force MAE at a performance level of 0.055 eV/\AA, the validation mean force MAE shows no further downward trend, and 1\% of the training configurations and 0\% of the validation configurations possess a force MAE above the 0.1 eV/\AA\ threshold.

The MAE in the MLIP energy predictions relative to the DFT ground truth follow a similar trend to the forces (Figure~\ref{fgr:3}C,D), with an uptick in the error as the configurational diversity increases in Round 2, followed by a precipitous decrease and plateau in the training and validation mean energy MAE around 0.001 eV/atom in Rounds 3-6. State-of-the-art MLIPs typically achieve energy MAEs in the range 0.001-0.005 eV/atom \cite{Zhang2024b}. By the terminal Round 6 of the active learning campaign and after having been exposed to just 1800 training configurations, fully 100\% of training configurations and 100\% of validation configurations possess an energy MAE better than 0.005 eV/atom. 

Taken together, these results demonstrate that the trained SiC Allegro MLIP has force and energy prediction accuracy at at a level of \textit{ab initio} DFT calculations and a diversity-maximizing active learning strategy enabled training of the model in a computationally and data efficient manner.

\subsection{Prediction of defect transition free energy barriers} 

\begin{figure}[ht!]
  \includegraphics[width=\textwidth]{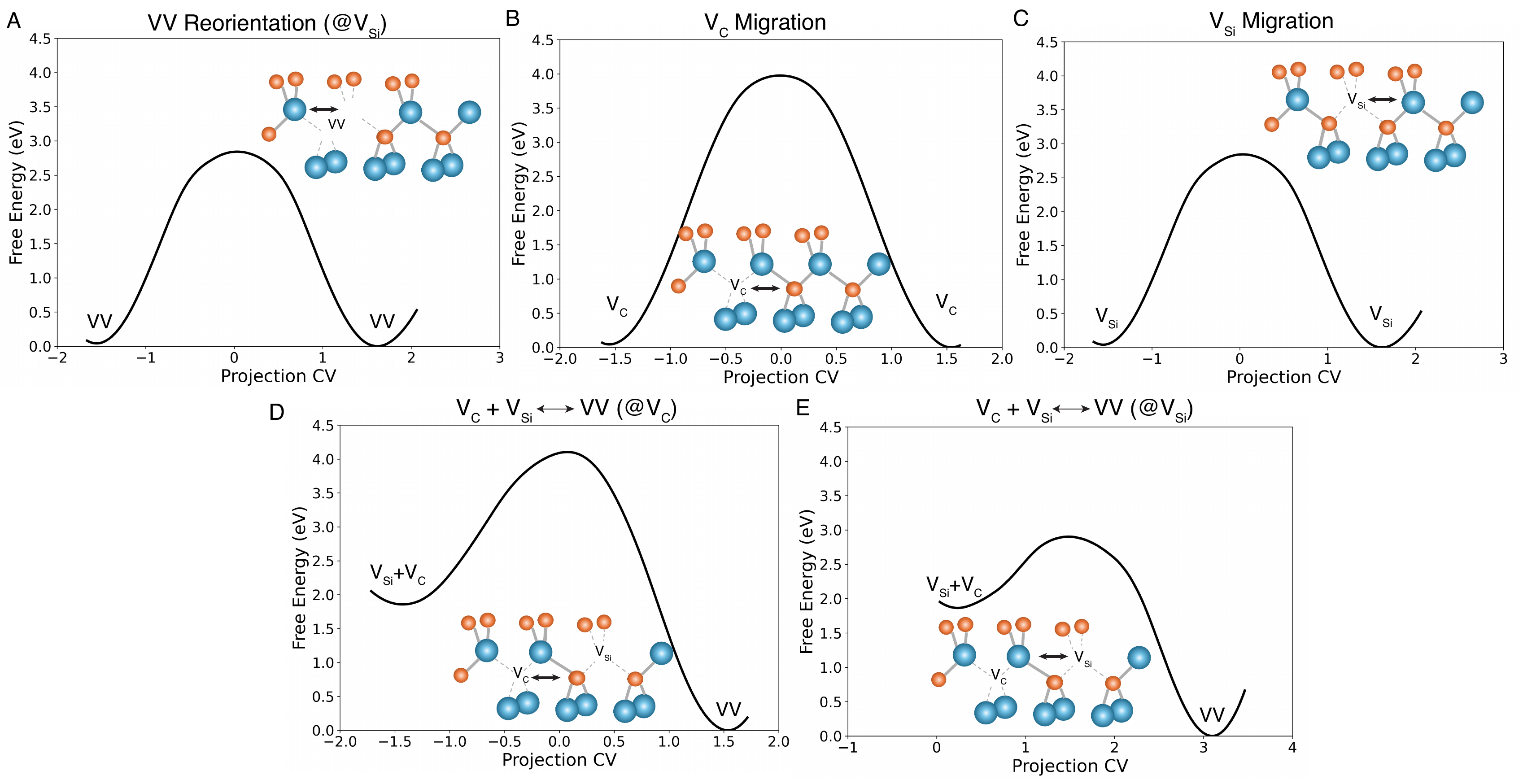}
  \caption{Free energy landscapes for five defect transition processes estimated under the trained Allegro MLIP using ABF: (A) reorientation of a VV divacancy via motion of a Si atom, (B) migration of a V$_{C}$ monovacancy, (C) migration of a V$_{Si}$ monovacancy, (D) VV divacancy formation from a pair of monovacancies via C-atom hopping, and (E) VV divacancy formation from a pair of monovacancies via Si-atom hopping. In each case, the free energy landscape is projected into one dimensional projection CV as described by Zhang \textit{et al.} \cite{Zhang2024}.}
  \label{fgr:4}
\end{figure}

We next evaluated the performance of the trained Allegro MLIP in predicting the activation free energies for a variety of known important defect transition processes in SiC \cite{Zhang2024}. This is a quite stringent test for MLIP performance as it requires correct prediction of the dynamical transition mechanism, the transition state, and the associated energy. Five different defect transitions were considered: (i) reorientation of a VV divacancy via motion of a Si atom, (ii) migration of a V$_{C}$ monovacancy, (iii) migration of a V$_{Si}$ monovacancy, (iv) VV divacancy formation from a pair of monovacancies via C-atom hopping, and (v) VV divacancy formation from a pair of monovacancies via Si-atom hopping. The first three transitions are symmetric in the sense that the initial and final states represent equivalent configurations and so possess identical free energies, whereas the final two transitions are asymmetric, with the forward process representing divacancy formation and the reverse process divacancy annihilation. We present in Figure~\ref{fgr:4} a schematic illustration of each defect transition process along with the free energy landscapes for each process computed at 1500 K under the terminal trained Allegro MLIP using the the adaptive biasing force (ABF) method \cite{Darve2008} implemented within the SSAGES software framework \cite{Sidky2018}. Details of the ABF implementation are provided in the \blauw{Supplementary Methods} in the \blauw{Supporting Information}. In each case, the ABF calculations were carried out using a one-dimensional collective variable (CV), defined as the projection of the distance between the moving atom and the center of mass of a selected group of atoms onto the vector connecting the atom’s initial and final positions, where the choice of this CV and the corresponding atom selection for each process follow the methodology described by Zhang \textit{et al.} \cite{Zhang2024}.

We present in Table~\ref{tab:1} a quantitative comparison of the Allegro MLIP activation free energies to the corresponding values computed using \textit{ab initio} molecular dynamics calculations with the ABF method. The \textit{ab initio} molecular dynamics free energy barriers for VV reorientation and formation/annihilation were taken from prior work by Zhang \textit{et al.}\ \cite{Zhang2024}, and those for $V_C$ and $V_{Si}$ migration computed in this work using an identical protocol. All \textit{ab initio} molecular dynamics calculations were performed in Qbox \cite{Gygi2008} at 1500 K using the SSAGES \cite{Sidky2018} implementation of ABF \cite{Darve2008}. Full details of the calculations are provided in the \blauw{Supplementary Methods}. In all cases the Allegro MLIP free energy barriers are in agreement with the \textit{ab initio} values within thermal fluctuations (1 $k_BT$ $\approx$ 0.13 eV at 1500 K), although we do note that in all cases the MLIP value tends towards higher values than the \textit{ab initio} result.

\begin{table}[ht!]
\centering
\caption{Comparison of the free energy barriers for the five defect transition processes illustrated in Fig.~\ref{fgr:4} computed at 1500 K using \textit{ab initio} molecular dynamics and molecular dynamics under the trained Allegro MLIP. Free energy landscapes were computed using the adaptive biasing force (ABF) method. For symmetric transitions, a single value for the barrier height is reported; for asymmetric transitions, the forward and backward barrier heights are reported. The final column reports the difference $\Delta$ between the MLIP and \textit{ab initio} free energy barrier values. For reference, at 1500 K 1 $k_BT$ $\approx$ 0.13 eV.}
\begin{tabular}{|l|c|c|c|}
\hline
\textbf{Defect Transformation} & \textbf{\textit{ab initio} (eV)} & \textbf{MLIP (eV)} & \textbf{$\Delta$ (eV)} \\
(T = 1500 K) & Forward (Backward) & Forward (Backward) &  \\
\hline
VV reorientation (@V$_\text{Si}$) & 2.70 & 2.82 & 0.12 \\
\hline
V$_\text{C}$ migration & 3.92 & 3.95 & 0.03 \\
\hline
V$_\text{Si}$ migration & 2.95 & 2.98 & 0.03 \\
\hline
V$_\text{C}$ + V$_\text{Si}$ $\rightarrow$ VV@V$_\text{C}$ & 2.24 (4.06) & 2.25 (4.10) & 0.01 (0.04) \\
\hline
V$_\text{C}$ + V$_\text{Si}$ $\rightarrow$ VV@V$_\text{Si}$ & 1.14 (2.98) & 1.04 (2.90) & 0.10 (0.08) \\
\hline
\end{tabular}
\label{tab:1}
\end{table}

The excellent accuracy of the trained Allegro MLIP in predicting atomic energies and forces and defect transition free energy barriers relative to ground truth DFT and \textit{ab initio} molecular dynamics calculations provides a strong validation of the accuracy and reliability of the trained MLIP. These results support the utility of the MLIP as a significantly more computationally efficient alternative to \textit{ab initio} molecular dynamics for the study of defect formation and transitions in SiC crystals.

\subsection{Transferability of trained MLIP to multi-defect systems} 

As a next test of the trained Allegro MLIP, we sought to assess its transferability to multi-defect systems despite having been trained exclusively on configurations containing divacancies and monovacancy pairs. This analysis tests the degree to which the local structural environments the model was exposed to in the divacancy and monovacancy pair training configurations are sufficiently representative of those encountered in multi-defect systems to permit robust, accurate, and transferable force predictions. 

\begin{figure}[!ht]
  \includegraphics[width=0.5\textwidth]{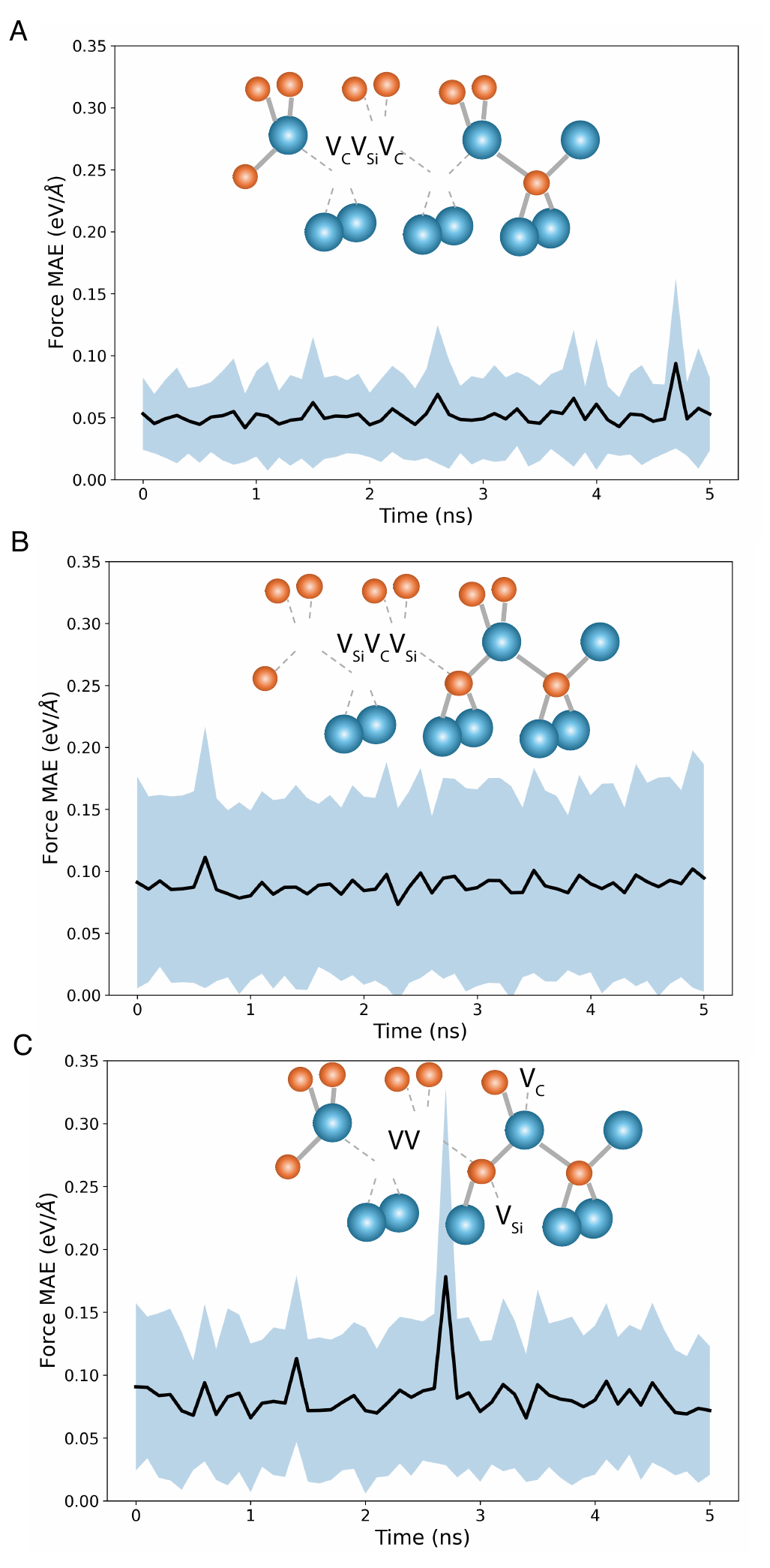}
  \caption{Force accuracy of trained Allegro MLIP in extrapolative predictions to (A) $V_C V_{Si} V_C$ and (B) $V_{Si}$V$_C$V$_{Si}$ trivacancy defect simulations and (C) multivacacancy defect simulations. The MAE of the MLIP force predictions relative to the DFT calculations are illustrated over the simulation time series. The black lines and shaded regions represent the MAE and the standard deviation of the per-atom force error, respectively, computed at 100 ps time intervals.} 
  \label{fgr:5}
\end{figure}

First, we constructed two trivacancy systems within a 216-atom 3C-SiC simulation box -- one containing two carbon and one silicon vacancy ($V_C V_{Si} V_C$) and one containing one carbon and two silicon vacancies ($V_{Si} V_C V_{Si}$) -- and conducted 5 ns of unbiased molecular dynamics simulations under the trained Allegro MLIP in each of the two systems at 1500 K. To assess the accuracy of the MLIP force predictions, we extracted structures at 100 ps intervals and, using the same approach used to assign ground truth forces and energies to the MLIP training data, computed for comparison the forces using DFT under the PBE functional. As illustrated in Figure~\ref{fgr:5}A and B, in each system the mean force MAE remains below approximately 0.10 eV/$\mathrm{\AA}$, corresponding to the accuracy threshold at which the models were trained. Moreover, this error remained flat and did not show any increase over the course of the simulation that may be indicative of an inherent instability in the MLIP.

Second, we constructed a larger 512-atom 3C-SiC simulation cell and introduced ten randomly distributed vacancy defects by removing ten atoms. Specifically, six carbon and four silicon atoms were removed, resulting in one divacancy, five carbon vacancies, and three silicon vacancies. We conducted 5 ns of molecular dynamics simulation at 1500 K under the Allegro MLIP, harvested configurations at 100 ps intervals, and computed the force MAE relative to DFT force calculations under the PBE functional. As illustrated in Figure~\ref{fgr:5}C, the mean force MAE remained below approximately 0.10 eV/\AA\, with the exception of a transient excursion to 0.15 eV/\AA\ at around 2.7 ns.

These analyses demonstrate that the trained Allegro MLIP retains high predictive accuracy in applications to more complex multi-defect configurations to which it was not exposed during training. Not only does the MLIP enable the simulation of far larger systems for far longer time scales than would typically be accessible to \textit{ab initio} calculations, but the simulations remain stable over the course of a 5 ns run and the predictive accuracy of the forces show little to no significant degradation relative to the 0.10 ev/\AA\ accuracy threshold to which the model was trained. 

\subsection{Prediction of temperature effects}

The MLIP was trained on DFT forces and energies collected from a variety of configurations generated by enhanced sampling. As such, the model was not parameterized at any particular temperature and, provided the configurations encountered at any temperature of interest were adequately represented within the enhanced sampling training data, the model is expected to be temperature transferable. Of particular interest in regards to temperature effects, is to predict how temperature influences the kinetics and thermodynamics of divacancy formation and to identify temperature annealing protocols to selectively promote and stabilize divacancies as putative qubits for quantum information technologies. 

We performed 250 nanoseconds of unbiased simulation of a 216-atom SiC box containing a single divacancy at five different temperatures: 1000 K, 1250 K, 1500 K, 1750 K, and 2000 K. To characterize the kinetics and thermodynamics of defect transitions at each temperature, we employed a Markov State Modeling (MSM) framework \cite{Prinz2011} using the PyEMMA software \cite{Scherer2015}. In brief, we generated 250 ns of unbiased simulation trajectories at each temperature by employing 1000 independent 250 ps simulations initialized from diverse initial configurations, projected these data into a 3D latent space spanned by the three leading slowest dynamical modes of the 1000 K simulations, performed a microstate clustering using k-means into a microstate MSM with a lag time of 45 ps, followed by a macrostate clustering using the Perron Cluster Cluster Analysis (PCCA+) method \cite{Rblitz2013} to generate a macrostate MSM for each temperature. The macrostate MSMs can be viewed as interpretable surrogate models with which to understand, predict, and control defect kinetics and thermodynamics. Full details of these calculations are provided in the \blauw{Supplementary Methods}.

\begin{figure}[!ht]
  \includegraphics[width=\textwidth]{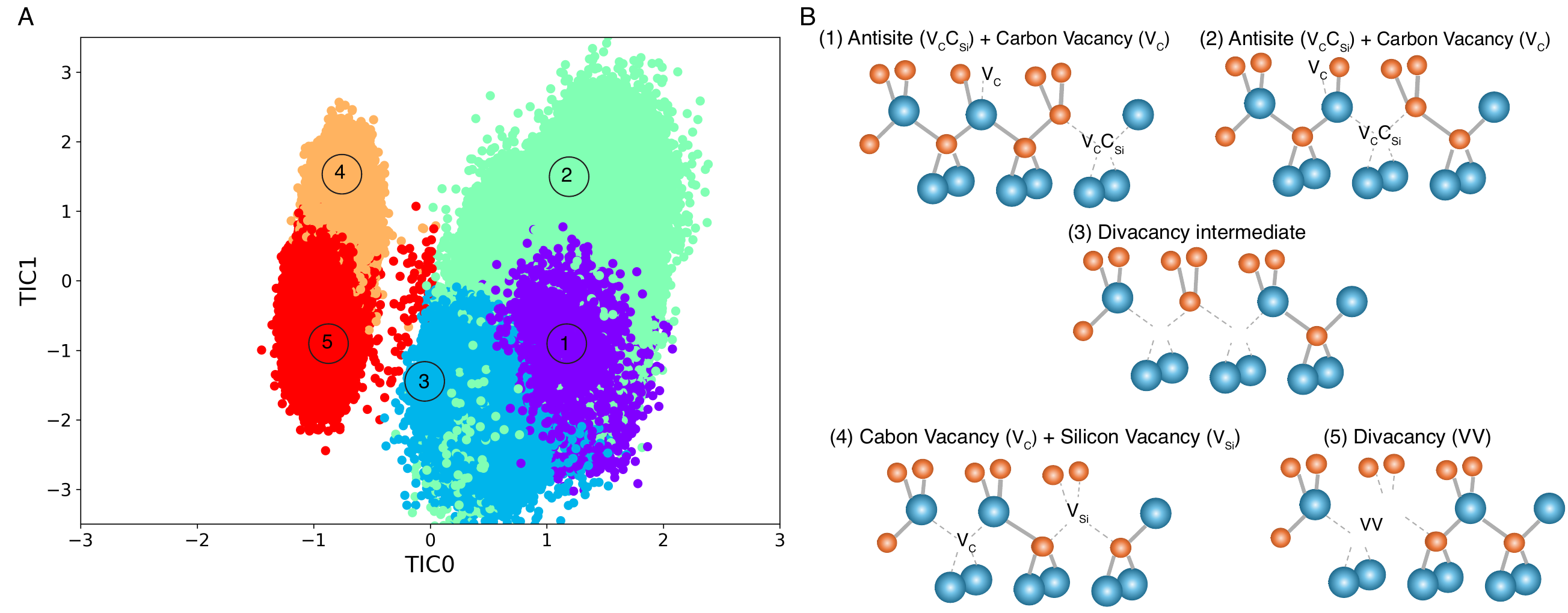}
  \caption{Structural interpretation of the macrostates within the 1500 K macrostate MSM. (A) Projection of the 1500 K simulation data into the leading two slowest dynamical modes of the system dynamics TIC0-TIC1 identified by applying TICA to the 1000 K simulation trajectories. Macrostates were defined by k-means clustering within a 3D TICA projection to form a microstate MSM, then PCCA+ clustering of the microstate MSM into a macrostate MSM. The projected configurations corresponding to each of the five macrostates is indicated by a different color and numerical label. (B) Inspection of the constituent configurations reveals each macrostate to correspond to a particular defect environment comprising a variety of structurally similar configurations. Schematic representations of the defect environment corresponding to each macrostate.}
  \label{fgr:6}
\end{figure}

Analysis of the macrostate MSMs at each temperature revealed each macrostate to clearly correspond to one of five defect environments: monovacancies, divacancy, divacancy intermediate, and two antisite defects. We illustrate in Figure~\ref{fgr:6} the structural interpretation of the five macrostates for the 1500 K model. Similar trends were observed for the MSMs at all five temperatures indicating that the thermally accessible configurations were well conserved across the 1000-2000 K temperature window. We did not observe the appearance of any new defect configurations that were not observed in the training data, suggesting that the diversity maximizing strategy employed in the active learning cycle used to parameterize the model most likely explored all of the thermally-accessible defect states.

\begin{figure}[!ht]
  \includegraphics[width=\textwidth]{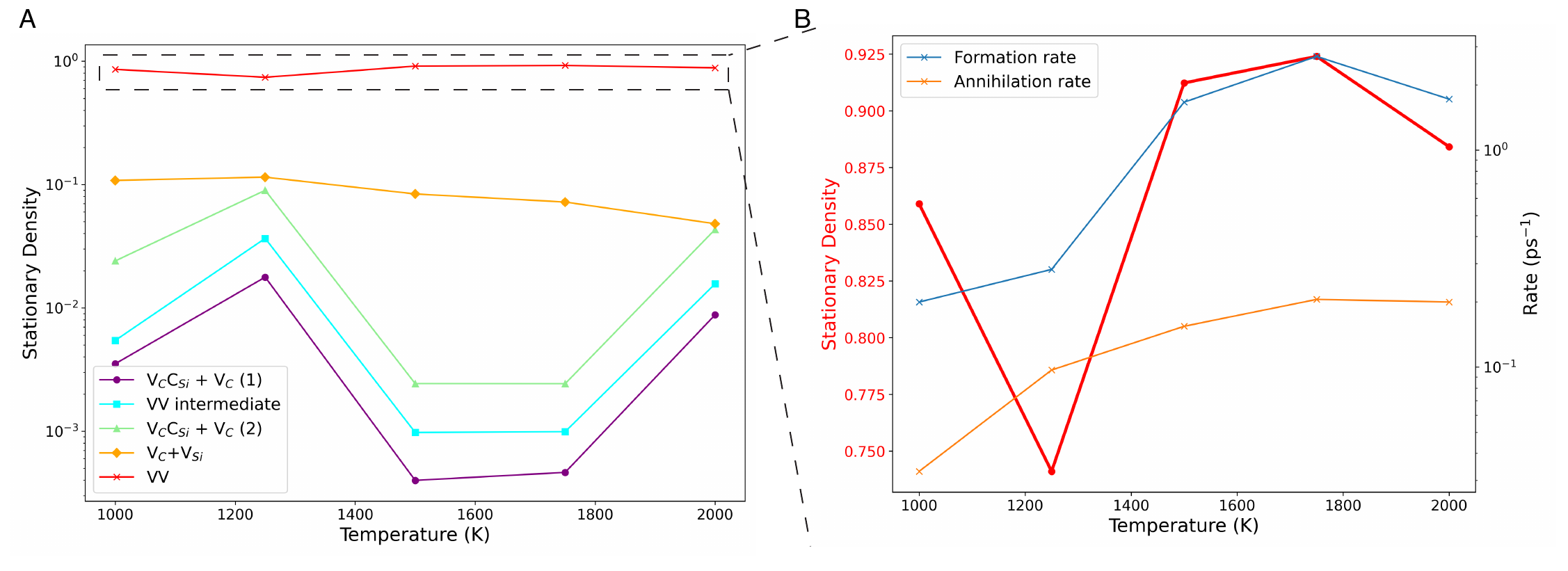}
  \caption{Thermodynamic stability of defect macrostates and kinetics of interconversion as a function of temperature exposed by the macrostate MSMs. (A) Stationary densities (i.e., equilibrium probability distribution) of the five defect macrostates at each temperature computed from the macrostate MSM. (B) Temperature dependence of the divacancy stationary density (red; left axis) and its formation and annihilation rates (blue and orange, respectively; right axis) computed by transition path theory (TPT) applied to the macrostate MSMs at each temperature.}
  \label{fgr:7}
\end{figure}

We next analyzed how the relative stabilities of, and transition rates between, the macrostates varied with temperature. First considering thermodynamic stability, we present in Figure~\ref{fgr:7}A the stationary densities within each macrostate corresponding to the equilibrium probability distribution. We observe the VV divacancy to be the most stable defect configuration at all temperatures in the range, but that its stability varied in a non-monotonic fashion. In particular, the divacancy stationary density decreased between 1000 K and 1250 K, then increased again to a maximum at 1750 K. To understand this behavior, we estimated the rates of divacancy formation and annihilation by applying transition path theory (TPT) to the macrostate MSM to compute the net flux into (formation) and out of (annihilation) the divacancy macrostate by all possible pathways \cite{No2009}. TPT reveals the divacancy formation and annihilation rates to both increase with temperature, but that the between 1000 K and 1250 K, the annihilation rate grew relatively faster than the formation rate, leading to an overall destabilization of the divacancy macrostate at the higher temperature and a $\sim$14\% relative reduction in its equilibrium stationary density (Figure~\ref{fgr:7}B). As required by overall conservation of probability, this drop coincided with corresponding increases in other defect populations: specifically the concentrations of the divacancy intermediates and antisite defects increased by approximately sevenfold and fourfold, respectively. As the temperature is increased beyond 1250 K, the divacancy stationary density rises again, peaking at 1750 K. Again, this trend can be understood from the temperature dependence of the formation and annihilation rates, wherein the former increases more rapidly than the latter over this range. This result is also consistent with the divacancy activation temperature previously predicted by first principles molecular dynamics \cite{Zhang2024}. Finally, at the highest 2000 K temperature, the divacancy equilibrium probability falls off as the formation rate decreases slightly while the annihilation rate remains approximately constant. 

The divacancy stability trends exposed by the macrostate MSMs predict the existence of a temperature optimum at which divacancy defects are maximally stable within the 1000-2000 K temperature window. Since MSMs were only constructed at five discrete temperatures in this range, we fitted a cubic spline to the temperature-dependent divacancy stability to predict a temperature of maximum stability of approximately 1625 K (Figure~S14). 

A number of prior experimental studies have considered the thermal stability and annealing behavior of VV defects, primarily in the 4H-SiC polytype. Experimental work by Carlsson \textit{et al.}\ probed VV concentrations as a function of annealing temperature using electron paramagnetic resonance (EPR) spectroscopy to observe an approximately monotonic increase in VV spin defect concentration between 273 K and 1273 K \cite{Carlsson2010}. A study by Karsthof \textit{et al.}\ employing photoluminescence (PL) spectroscopy came to a similar conclusion, with the VV-related PL signal monotonically increasing over the 573 K to 1273 K temperature range 
\cite{karsthof2020conversion}. Son \textit{et al.}\ conducted EPR studies of annealing temperatures over the range 873 K to 1873 K, and identified a peak in the VV concentration at around 1773 K \cite{Son2006, Son2007}. A higher-temperature EPR study by Carlos \textit{et al.}\ measured a monotonic decrease in the VV concentration over a temperature range of 1673 K to 2073 K \cite{Carlos2006}. A recent study by Kimura \textit{et al.}\ employed PL spectroscopy to measure the VV concentration at annealing temperatures of 1123 K, 1273 K, and 1473 K, with the 1273 K annealing temperature producing the highest VV-related PL intensity \cite{Kimura2025}. By appealing to \textit{ab initio} calculations \cite{Zhang2023, Zhang2024}, these observations were understood as the lower 1123 K annealing temperature as offering insufficient thermal energy to activate silicon vacancy (V$_{Si}$) migration necessary for aggregation with a carbon vacancy (V$_C$) to form the VV divacancy, whereas the higher 1473 K annealing temperature offered sufficient thermal energy to also mobilize the VV and V$_C$ defects \cite{karsthof2020conversion, bathen2019anisotropic}. The 1273 K annealing temperature is optimal in providing sufficient thermal energy for V$_{Si}$ migration for VV formation, but insufficient thermal energy to activate VV migration and permit their migration into deeper regions of the crystal or towards the surface where they are annihilated. Since the VV divacancy is known to be stable up to much higher temperatures \cite{Carlos2006}, the decrease in the measured VV-related PL intensity is therefore predicted to be due to migration and/or annihilation of the VV divacancies as opposed to their intrinsic dissociation. 

In comparing the annealing temperature predictions of our model to these experimental data, we must take care to view this through a number of caveats associated with the limitations of our model and simulation protocols. First, our simulations consider the 3C-SiC polytype, whereas the experiments pertain to the 4H-SiC polytype. Second, our simulations consider a 216-atom supercell with periodic boundaries containing a single divacancy. As such, our calculations effectively model a homogenous bulk crystal at a specified defect concentration. The defect can and does undergo transitions between the five defect configurations -- monovacancies, divacancy, divacancy intermediate, and two antisite defects (Figure~\ref{fgr:6}) -- and the defects are also free to migrate. However, our simulation setup precludes migration-driven loss of experimentally observable VV divacancies through diffusion into the crystal interior where they may evade PL spectroscopy or migration to the crystal surface where they may undergo annihilation. Moreover, our calculations on a single divacancy also preclude the opportunity for more complex multi-defect configurations and dynamics. Third, our MLIP was parameterized for a single charge state (neutral) and spin state (triplet) that remain fixed over the course of the simulation. This is the most significant deficiency of the model and the most likely source of discrepancies with measurements of experimental systems that can, of course, undergo changes in charge and spin state over the course of their dynamical evolution. 

With these caveats in mind, our predictions are consistent with the experimental work of Son \textit{et al.}\ in that we predict an optimum temperature for maximization of the VV concentration, and our prediction of the maximum at around 1625 K is only 150 K lower than the 1773 K maximum in the measured VV-associated EPR signal \cite{Son2006, Son2007}. Our prediction is consistent with the EPR measurements of Carlos \textit{et al.}, who report a monotonic decrease in the VV-associated signal from 1673 K to 2073 K, but this study does not report the lower temperature data required to test our prediction of a maximum at approximately 1625 K \cite{Carlos2006}. Given the inherent insufficiency of our 216-atom simulation cell in modeling VV migration into the deep crystal or surface annihilation, we are wary to draw too close a comparison with the PL measurements of Kimura \textit{et al.}\ given the centrality of these mechanisms in rationalizing the experimental observations \cite{Kimura2025}. Our calculations effectively isolate the intrinsic thermodynamic stability of a single divacancy, unaffected by extrinsic processes such as migration or surface annihiliation. In this more restricted sense, however, we draw a point of consistency with Kimura \textit{et al.}\ in that our calculations predict the VV divacancy to be intrinsically stable up to approximately 1650 K, supporting the mechanism put forward in that study that the reduction in the observed VV-associated PL intensity beyond 1273 K is not due to VV dissociation but rather migration into the crystal interior or surface \cite{Kimura2025}.

\section{Conclusions}

In this study, we developed an E(3)-equivariant Allegro MLIP to model defect dynamics in the 3C-SiC crystal polytype. The trained model possesses \textit{ab initio} accuracy in predicting atomic energies and forces, while possessing computational efficiencies approaching classical force fields that permit the efficient generation of long trajectories for large systems that are outside the reach of \textit{ab initio} molecular dynamics. We devise an efficient active learning framework to accumulate training configurations for the model by integrating data-driven learning of symmetry-adapted collective variables, enhanced sampling of the thermally-accessible phase space, a physically-motivated prior potential, and a diversity maximizing selection of training configurations. Using this strategy, we converged the Allegro MLIP to a force prediction accuracy of 0.1 eV/\AA\ using just 1800 training configurations over the course of six active learning rounds, and deploying a total of $\sim$240 GPU-hours on MLIP molecular dynamics simulations, $\sim$72,000 CPU-h in computing ground truth DFT energies and forces, and $\sim$43 GPU-h in MLIP training. We demonstrated excellent agreement of the MLIP predictions of defect transition free energy barriers to \textit{ab initio} calculations, verified the transferability and stability of the trained MLIP to larger, multi-defect systems, and conducted an analysis of the temperature dependence of defect thermodynamic stability to propose an optimal annealing temperature to maximally stabilize VV divacancies within the SiC lattice and an analysis of the formation/annihilation kinetics to provide atomic-level dynamical insight into the stability mechanism.

This study makes two key contributions. First, it demonstrates a generalizable active learning framework for the data and compute efficient training of MLIPs with \textit{ab initio} accuracy in atomic energy and force prediction. Second, it delivers an accurate and validated MLIP for 3C-SiC force field that can be extended to simulate, understand, and control the thermodynamics and dynamics of lattice defects. This study illustrates the value of MLIPs as an enabling tool to simulate defect dynamics at far larger time and length scales than are accessible to \textit{ab initio} calculations to expose atomic-level understanding of the defect formation, migration, and stabilization mechanisms critical to defect engineering efforts. The simulation protocols presented in this work are also of utility for the efficient training of MLIPs in condensed matter systems, and their deployment in predicting important materials properties such as free energy barriers, optimal annealing temperatures, and defect transition mechanisms as emergent phenomena within unbiased molecular dynamics simulations, through the parameterization of MSMs as long-time kinetic models, and via coupling to efficient collective variable biasing and path-based enhanced sampling techniques. 

The good agreement with experimental measurements of the temperature stability of VV divacancies and prior first principles predictions of activation energies is encouraging, but we also identify a number of avenues for additional model development. First, the present MLIP development assumes fixed charge and spin states for all configurations. In reality, these properties can change during defect transformations and strongly influence defect formation and migration energetics \cite{Zhang2023, Zhang2024}. While it would be straightforward to train an ensemble of MLIPs at different charge and spin states using the approach presented herein, the incorporation of variable charge and spin into the MLIP using ideas from recent developments in charge- and spin-aware potentials \cite{Yuan2024, ueno2024spinmultinetneuralnetworkpotential, Deng2023, Unke2021, Kalita2025} would significantly enhance the physical realism of the model. Second, while this work focuses on a single polytype of SiC, the material exists in multiple polytypes with distinct structural and electronic properties. Developing a transferable, universal MLIP that accurately describes defects across all major SiC polytypes would greatly broaden the applicability of the model. Third, we have verified that the trained MLIP produces stable and accurate simulation trajectories in applications to larger multi-defect systems, but deploying an analogous active learning loop employing larger and more diverse systems would be of value in further fine-tuning and improving the predictive accuracy of the model for more complex multi-defect configurations and transitions.

\section*{Author Contributions}

A.L.F., G.G., and F.G.\ designed the research. S.D.\ designed and executed the active learning workflow for MLIP parameterization, including collective variable discovery and enhanced sampling, DFT energy and force calculations, training of the MLIP model, and convergence assessment. C.Z. and S.D.\ computed the defect transition free energy barriers. G.P.L.\ provided guidance in MLIP training and evaluation. A.L.F., G.G., F.G., and J.J.dP.\ supervised the research. S.D.\ and A.L.F. wrote the manuscript. A.L.F., G.G., F.G., and C.Z.\ reviewed and critically revised the manuscript. All authors read and approved the final manuscript

\section*{Acknowledgements}

This work was supported by MICCoM (Midwest Center for Computational Materials), as part of the Computational Materials Science Program funded by the U.S. Department of Energy, Office of Science, Basic Energy Sciences, Materials Sciences and Engineering Division, through Argonne National Laboratory, under contract no.~DE-AC02-06CH11357. This work was completed in part with resources provided by the University of Chicago Research Computing Center. We gratefully acknowledge the computing time at the University of Chicago high-performance GPU-based cyber infrastructure supported by the National Science Foundation under grant no.\ DMR-1828629.

\section*{Conflict of Interest Disclosure}

A.L.F.\ is a co-founder and consultant of Evozyne, Inc.\ and a co-author of US Patent Applications 16/887,710 and 17/642,582, US Provisional Patent Applications 62/853,919, 62/900,420, 63/314,898, 63/479,378, 63/521,617, and 63/669,836, and International Patent Applications PCT/US2020/035206, PCT/US2020/050466, and PCT/US24/10805.

\section*{Code and Data Availability}

The codes implementing the active learning workflow for training the Allegro MLIP -- including PINES CV discovery, OPES-Explore enhanced sampling, SOAP dissimilarity training point selection, DFT calculation of forces and energies, convergence assessment, and Allegro MLIP training -- along with the trained Allegro MLIP are made available via a GitHub repository at \url{https://github.com/Ferg-Lab/SiC_Allegro_MLIP.git}. 

\clearpage
\newpage

\bibliography{manuscript}


\end{document}